\documentclass[10pt, final,  doublecolumn]{IEEEtran}
\usepackage{cite}
\usepackage{algorithm2e}
\usepackage{amssymb}
\usepackage{enumerate}
\usepackage{graphicx}
\usepackage{amssymb}
\usepackage{amsbsy}
\usepackage{float}
\usepackage{balance}
\usepackage{url}
\usepackage{bm}
\usepackage{array}
\usepackage{varioref}
\usepackage[normalem]{ulem}
\usepackage{color}
\usepackage{empheq}
\usepackage{flushend}
\begin{document}

\bstctlcite{asms:BSTcontrol}
\title{Frame Based Precoding in Satellite Communications: A Multicast  Approach}
\author{\IEEEauthorblockN{ Dimitrios Christopoulos\IEEEauthorrefmark{1},   Symeon Chatzinotas\IEEEauthorrefmark{1}
 and Bj\"{o}rn Ottersten\IEEEauthorrefmark{1}\\
 }

\IEEEauthorblockA{\IEEEauthorrefmark{1}SnT - securityandtrust.lu,  University of Luxembourg
\\email: \textbraceleft dimitrios.christopoulos, symeon.chatzinotas, bjorn.ottersten\textbraceright@uni.lu}

%
}
\maketitle


\begin{abstract}
In the present work, a multibeam satellite that employs aggressive frequency reuse towards increasing the offered throughput is considered. Focusing on the forward link, the goal is to employ multi-antenna signal  processing techniques, namely linear precoding, to manage the inter-beam  interferences. In this context, fundamental practical limitations, namely the rigid   framing structure of satellite communication standards and the on-board per-antenna power constraints, are herein considered. Therefore, the concept of  optimal frame based precoding under per-antenna constraints, is discussed. This  consists in precoding the transmit signals without changing the underlying framing structure of the communication standard.  In the present work, the connection of the frame based precoding problem with the generic signal processing problem of conveying independent sets of common data to distinct groups of  users is established.   This model is known as physical layer multicasting to multiple co-channel groups. Building on recent results, the weighted fair per-antenna power constrained multigroup multicast precoders are employed for  frame based precoding.   The throughput performance of these solutions is compared to multicast aware heuristic precoding methods over a realistic multibeam satellite scenario. Consequently, the gains of the proposed approach are quantified via extensive numerical results.
\end{abstract}
 \begin{IEEEkeywords}
Frame based Precoding; Physical layer Multigroup Multicasting;    Per-antenna Power Constraints; Multicast Aware MMSE;  Weighted Max Min Fair Optimization;
\end{IEEEkeywords}
 \section{Introduction \& Related Work\ }
  The spatial degrees of freedom offered by the multibeam satellite antenna are a valuable resource towards efficiently reusing the user link bandwidth. Advanced transmit signal processing techniques, namely beamforming (or equivalently precoding), are currently  employed to optimize the performance of the multi-antenna transmitters without compromising the complexity of single antenna receivers. This allows for  more aggressive frequency reuse schemes,  towards the   efficient utilization of the available spectrum and thus the  increase of the overall system throughput. As always, however, the benefits of these advanced schemes come with a cost. The most inhibiting  requisite for the application of linear precoding is the knowledge of the vector channel state information ($\mathrm{CSI}$) at the transmitter. Assuming readily available $\mathrm{CSI}$ at the transmit side,   full-frequency re-use schemes are foreseen to boost the throughput of the next generation broadband interactive high throughput multibeam satellite systems, by the means of linear precoding  \cite{Christopoulos2013AIAA}.

 Conventionally, the channel capacity achieving  precoding design assumes independent symbols, each  addressed to a different single user. This allows for a channel by channel calculation of the precoding matrices. However, such an assumption cannot apply in a satellite scenario. The framing structure of optimized with the inherent attributes of the satellite channel of legacy communication standards, imposes specific constraints in the practical implementation of precoding. The physical layer design of $\mathrm{DVB-S2}$  \cite{DVB_S2_standard} and $\mathrm{DVB-S2X}$ \cite{DVB_S2X}, encompasses long forward error correction codes ($\mathrm{FEC}$) to cope with noise limited channels and long propagation delays. Therefore, as the framing of multiple users per transmission is  emanated  to guarantee scheduling efficiency, the particularly long $\mathrm{FEC}$ frames inhibit precoding. Consequently, the traditional assumption of a single user terminal ($\mathrm{UT}$) per transmission is alleviated and the application of precoding methods in future satellite systems  relies on frame-by-frame precoding\cite{Christopoulos2013AIAA}.
\begin{figure*}[!t]
\centering
\includegraphics[width= 0.8\textwidth]{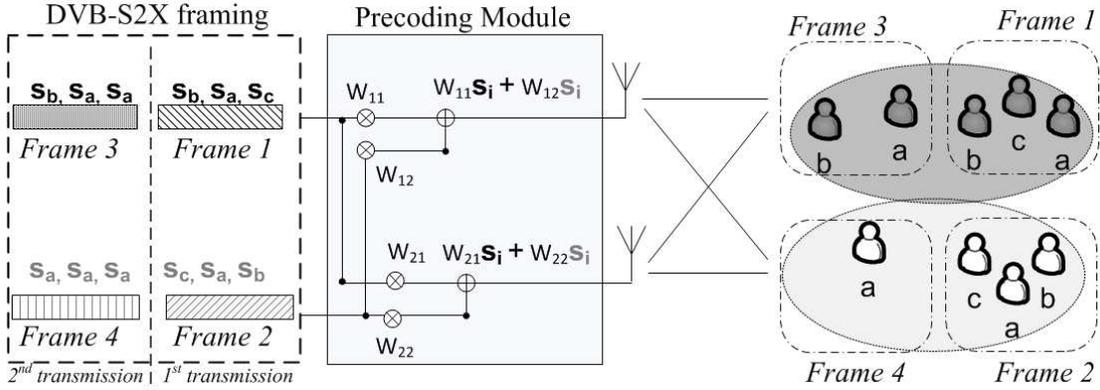}\\
\caption{Frame based precoding transmitter: a toy example. }
\label{fig: proposed system model}
\end{figure*}

  An toy example of the application of linear precoding in satcoms is given in Fig. 1. Therein four frames belonging to two subsequent transmissions are shown. Each transmitted frame addresses multiple users.  A symbol denoted as $s_{i}$ is addressed   to the $i$-th user of the corresponding frame. One should bear in mind that bit and symbol interleaving take place in a frame, while user data payloads are not always equal, as depicted in the second transmission in Fig. 1. In frames 3 and 4, different amount of data is transmitted to each user. It should be stressed that symbols denoted as $s_{a}$ are not carrying identical data.  To  simplify the analysis, in the following it will be assumed that in each beam, an equal number of $\mathrm{UT}$s  is co-scheduled in each frame. Hence, the first transmission instance of Fig. 1 will be modeled. Such an assumption can be realized in a practical system via the use of dummy data to fill frames.  Also, the frames are of constant size and transmissions amongst the beams are perfectly synchronized. These assumptions are in line with the latest evolution of DVB-S2X \cite{DVB_S2X} \cite{Morello2013AIAA}.

The purpose of the present work is to establish the connection of frame based precoding with the generic, physical layer multicasting problem \cite{Sidiropoulos2006}. More specifically, physical layer ($\mathrm{PHY}$) multicasting to multiple co-channel groups \cite{Karipidis2008} can provide the theoretically optimal precoders when a multi-antenna transmitter conveys independent sets of common data to distinct groups of users. This scenario is known as   multigroup multicasting. The connection of the frame based precoding problem with the $\mathrm{PHY}$ multigroup multicasting is clear under the following considerations.
Multicasting is based on the assumption that the same information is transmitted to multiple receivers. This leads to designing a precoding vector matched to more than one channel vectors. From a different perspective, the framing structure of communication standards imposes that the same precoder will apply over the symbols of more than one user belonging in the same frame. Since each frame is received and decoded by all co-group users, the design of an optimal  frame based precoder is given by solving a multicast multigroup optimization problem. Thus, multicasting allows for an analytically formal modeling of the problem. Therefore, in the context of frame based precoding, the fact that the same precoder needs to apply to the different data of many receivers due to the framing constraint, leads to a multicast consideration.

 The second practical constraint tackled herein involves the on-board per-antenna constraints imposed by conventional multibeam communication payloads. The lack of flexibility in sharing energy resources amongst the antennas of the transmitter is  in general a  common scenario in multi-antenna systems where individual amplifiers per  transmit antenna, are employed. Despite the fact that flexible amplifiers could be incorporated in some cases, specific communication systems cannot afford this design. Typical per-antenna power limited systems can be found in  multibeam satellite communications \cite{Christopoulos2013AIAA}, where flexible on board payloads are difficult to implement. Other examples can be found in distributed antenna systems, such as cooperative satellite constellations or swarms of cooperative nano-satellites.

The optimal transmit beamforming in the minimum total transmit power sense, assuming channel based precoding, was initially derived under sum power constraints $\mathrm{(SPC)}$ over all the transmit antennas  in \cite{Bengtsson2001,bengtsson1999}. Per-antenna transmit power constraints ($\mathrm{PAC}$s) were considered later on, in \cite{Yu2007}.
Moreover, the multiuser downlink beamforming problem in terms of maximizing the minimum $\mathrm{SNIR}$,  was optimally solved in \cite{Schubert2004}. The goal of the later  formulation is to increase the fairness of the system by boosting the  $\mathrm{SNIR}$ of the  user that is further away from a targeted performance. Hence,  the problem is commonly referred to as \textit{max--min fair}. Amid the extensive literature on multigroup multicast beamforming \cite{Karipidis2008}, the derivation of the optimal multigroup multicast precoders when a maximum limit is imposed on the  transmitted power of each antenna is non trivial.    In this direction, the weighted max--min fair multigroup multicast beamforming for a per-antenna power constrained  system has been derived in \cite{Christopoulos2014ICC,Christopoulos2014}.
 In the present work, this solution is applied in the context of multibeam satellite systems. This allows for the frame based precoding problem for satellite communications to be formulated in an analytically tractable manner.

The rest of the paper is structured as follows.   The description of the multigroup multicast satellite system model is given in Sec. \ref{sec: System Model}. Heuristic approaches to solve this issue are discussed in Sec. \ref{sec: heuristics}. The multicast multigroup optimization problem definition is described in Sec. \ref{sec: optimal}.  In Sec. \ref{sec: performance}, the performance of the proposed technique over the multibeam satellite system  is evaluated. Finally,  Sec. \ref{sec: conclusions} concludes the paper and paves the way forward.

{\textit{Notation}: In the remainder of this paper, bold face lower case and upper case characters denote column vectors  and matrices, respectively. The operators \(\left(\cdot\right)^\text{T}\), \(\left(\cdot\right)^\dag\), $|\cdot|$ and \(||\cdot||_2, \) correspond to   the transpose, the conjugate transpose,  the absolute value and the Frobenius norm operations over matrices and vectors,  while $[\cdot]_{ij}  $  denotes the $i, j$-th element of a matrix.  

\section{System Model}\label{sec: System Model}
 Let us focus on a multi-user ($\mathrm{MU}$) multiple input single output ($\mathrm{MISO}$) multicast system. Assuming a single transmitter, let   $N_t$ denote the number of transmitting elements  and  $N_{u}$ the  total number of users served.  The input-output analytical expression  will read as $y_{i}= \mathbf h^{\dag}_{i}\mathbf x+n_{i},$
where \(\mathbf h^{\dag}_{i}\) is a \(1 \times N_{t}\) vector composed of the channel coefficients (i.e. channel gains and phases) between the \(i\)-th user and the  \(N_{t}\) antennas of the transmitter, \(\mathbf x\) is the \(N_{t} \times 1\)  vector of the transmitted symbols and  \(n_{i}\) is the independent complex circular symmetric (c.c.s.) independent identically distributed (i.i.d) zero mean  Additive White Gaussian Noise ($\mathrm{AWGN}$)  measured at the \(i\)-th user's receive antenna.

By defining a single multicast group per beam, in each transmission,  the multigroup multicast model of \cite{Karipidis2008} is realized. Thus, a total of $ G = N_{t}$ multicast groups are assumed,  with  $\mathcal{I} = \{\mathcal{G}_1, \mathcal{G}_2, \dots  \mathcal{G}_G\}$ the collection of   index sets and $\mathcal{G}_k$ the set of users that belong to the $k$-th multicast group, $k \in \{1\dots G \}$. Each user belongs to only one group, thus $\mathcal{G}_i\cap\mathcal{G}_j=$\O ,$  \forall i,j \in \{1\cdots G\}$. Let $\mathbf w_k \in \mathbb{C}^{N_t \times 1}$ denote the precoding weight vector applied to the transmit antennas to beamform towards the $k$-th group.
Let us also denote the number of users per group as $\rho = N_u / G$. By collecting all user channels in one channel matrix, the general linear signal model in vector form reads as
\begin{equation}\label{eq: general input output}
 \mathbf y = \mathbf {H}\mathbf x + \mathbf n = \mathbf {H}\mathbf {Ws} + \mathbf{ n}
\end{equation}
where $ \mathbf {y \text{ and }  n } \in \mathcal{\mathbb{C}}^{N_{u}}$, $\mathbf {x} \in \mathbb{C}^{N_{t}}$ and $ \mathbf {H} \in \mathbb{C}^{N_{u} \times N_t}$.  The multigroup multicast scenario imposes  a precoding matrix $ \mathbf {W} \in \mathbb{C}^{N_t  \times N_t}$ that includes as many precoding vectors (i.e columns) as the number of groups ($ G = N_{t}$). This is the number of independent symbols transmitted, i.e. $\mathbf {s} \in \mathbb{C}^{N_{t}}$.
The assumption of independent information transmitted to different groups implies that the symbol streams $\{s_k\}_{k=1}^G$ are mutually uncorrelated and the total power radiated from the antenna array is equal to
\begin{align}
P_{tot} = \sum_{k=1}^ {N_t} \mathbf w_k^\dag \mathbf w_k= \mathrm{Tr\left( \mathbf {WW}^\dag\right)},
\end{align}
where $\mathbf {W}= [\mathbf w_1, \mathbf w_2, \dots\mathbf w_G].$
The power radiated by each
antenna element is  a  linear combination of all precoders and reads as
\cite{Yu2007}\begin{align}\label{eq: PAC}
P_n = \left[\sum_{k=1}^{N_t} \mathbf w_k \mathbf w_k^\dag \right]_{nn} =\left[ \mathbf {WW}^\dag\right]_{nn},
\end{align}
where $n \in \{1\dots  N_t\}$ is the antenna index.
The fundamental difference between the $\mathrm{SPC}$  of \cite{Karipidis2008} and the proposed $\mathrm{PAC}$ is clear in  \eqref{eq: PAC}, where instead of one,  $N_t$ constraints are realized, each one involving all the precoding vectors.

\subsection{Equivalent Channel Model} \label{sec: equivalent channel}

 Assuming an equal number of groups and antennas results to a  square precoding matrix, as seen in the previous section.  Therefore, an alternative, simplified channel model in the fashion of \cite{Christopoulos2013AIAA} and \cite{Taricco2014} can also be adopted towards providing a more tractable representation.    To facilitate the comprehension of system model, let us define multiple square channel matrices $\mathbf H_{[i]}$. The index $[i]$ corresponds to the different UTs per beam that need to be served by the same frame, i.e. $i = 1,\dots \rho$. Each matrix corresponds to a ``a single user-per-beam'' instance, which is the common assumption in satellite precoding literature (e.g. \cite{Christopoulos2012} and the references therein). To model the frame based precoding constraint, the general input-output signal model  can be defined as \cite{Christopoulos2013AIAA}:
\begin{equation}
 \mathbf y_{[i]} = \mathbf {H}_{[i]}\mathbf x_{[i]} + \mathbf n_{[i]} = \mathbf {H}_{[i]}\mathbf {Ws}_{[i]} + \mathbf n_{[i]}
\end{equation}
where $ \mathbf {y, x, n, s} \in \mathcal{C}^{N_{t}}$ , with $\mathcal{E}||\mathbf {n}||^2 = \sigma^2$ and $\mathcal{E}||\mathbf {s}||^2 = 1$, while $ \mathbf {H}_{[i]} \in \mathcal{C}^{N_t  \times N_t}$ is a one-user-per-beam instance of the total non-square channel matrix.  Also, since an equal number of antennas and groups is assumed,  $N_{u} = \rho\cdot N_t$. The above definition allows for the calculation of one \textit{equivalent} precoder $\mathbf W = f(\mathbf{H}_{[i]})$.
The function $ f $ can be chosen according to the design criteria \cite{Taricco2014}.

\subsection{Multibeam Satellite Channel }\label{sec: Satellite Channel}
The above general system model, is herein applied over a multibeam satellite channel explicitly defined as follows.
A 245 beam pattern that covers Europe is employed\cite{satnex}. The multibeam radiation pattern is depicted in Fig. \ref{fig: CA}.
 \begin{figure}[h]
 \centering
 \includegraphics[width=1\columnwidth]{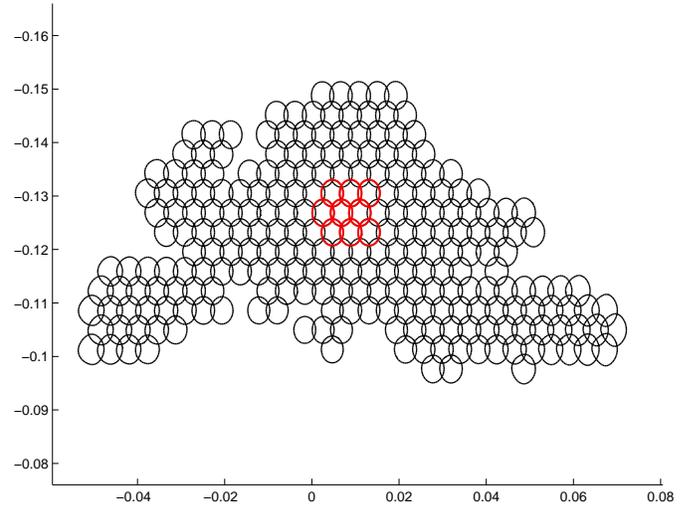}\\
  \caption{Plot of the coverage area with the 9 selected beams }
\label{fig: CA}
\end{figure}
A complex channel matrix that models the link budget of each UT as well as the phase rotations induced by the signal propagation and the payload is employed.
  In more detail, starting from the model followed in \cite{Zheng2012}, the total channel matrix    \( \mathbf{H}\in \mathbb{C}^{N_u \times N_t }\)
 is generated  as
\begin{equation}\label{eq: phase assumptions}
\mathbf{ H}=\mathbf{\Phi}_p\mathbf{B},
\end{equation}
and models the multibeam antenna pattern as well as the signal phase due to different propagation paths between the users.
The real matrix $\mathbf{B} \in \mathbb{R}^{N_u\times N_t}$ models the satellite antenna radiation pattern, the path loss, the receive antenna gain and  the noise power.  Its $i,j$-th entry is given by :
\begin{equation}\label{eq: beam_gain}
b_{ij}=\left(\frac{\sqrt{G_RG_{ij}}}{4\pi(d_k\cdot\lambda^{-1})\sqrt{\kappa T_{cs}W}}\right),
\end{equation}
with \(d_k\) the distance between the $k$-th UT and the satellite (slant-range), \(\lambda\) the wavelength, \(\kappa\) the Boltzman constant,  \(T_{cs}\) the clear sky noise temperature of the receiver, $W$ the user link bandwidth, $G_R$ the receiver antenna gain and $G_{ij}$ the multibeam antenna gain between the $i$-th single antenna UT and the $j$-th on board antenna (= feed). Hence, the beam gain for each antenna-UT pair, depends on the antenna pattern and on  the user position.
A fundamental assumption in multibeam satellite channels lies in assuming that one user will have the same phase between all transmit antennas due to the long propagation path \cite{Christopoulos2011,Christopoulos2012,Christopoulos2013,Zheng2012}. The identical phase assumption between one UT and all transmit feeds is supported by the relatively small distances between the transmit antennas and the long propagation distance of all signals to a specific receiver.  However,  this assumption discards any phase introduced by the on-board equipment due to imperfections and/or different on board propagation paths\footnote{More elaborate signal phase assumptions that consider the on-board payload imperfections, as well as phase offsets introduced by the receiver equipment, will be considered in future extensions of this work.}.  Hence, in \eqref{eq: phase assumptions} the diagonal square matrix $\mathbf \Phi_{p} \in \mathbb{C}^{N_u \times N_u }$ is generated as  $[\mathbf \Phi_{p}]_{kk} = e^{j\phi_k},\ \forall \ k = 1\dots N_{u} $ where $\phi_k$ is  a uniform random variable in $(0, 2\pi] $  and $[\mathbf \Phi_{p}]_{kn}=0, \ \forall \ k\neq n$.
\begin{table}
\caption{Link Budget \& Simulation Parameters}
\centering
\begin{tabular}{l|l}
\textbf{Parameter}  & \textbf{Value}\\\hline
 Frequency Band  &  K (20~GHz)   \\
 UT clear sky temp,   $T_{cs}$  & 235.3K\\
 User Link Bandwidth,   \(B_u\) & 500~MHz\\
 Output Back Off, $\mathrm{OBO}$&5~dB\\
 On board Power, $P_{tot}$ & 55~dBW\\
Roll off, $\alpha$ & 0.20 \\
 UT antenna  Gain,  $G_R$& 40.7~dBi \\
  Multibeam Antenna Gain, $G_{ij}$ &Ref: \  \cite{satnex}
  \\\hline
\end{tabular}
\label{tab: simulation params}
\end{table}

\section{Multigroup Multicast Precoding}

{  \subsection{Multicast Aware MMSE: Average Precoding}\label{sec: heuristics}
The optimal linear  precoder $\mathbf W = f(\mathbf H_{[i]} ), i=1\dots \rho$ in the  minimum mean square error sense,  with more users than transmit antennas is considered in this section. Under the constraint of designing a linear MMSE precoder $ \mathbf {W} \in \mathbb{C}^{N_t  \times N_t}$ for multiple channels, i.e. $\mathbf H \in \mathbb{C}^{N_{u}\times N_t}$ with ${N_{u} > N_t}$ the solution is not straightforward. Following the equivalent channel notation of Sec. \ref{sec: equivalent channel},   the problem of minimizing the MSE between the transmitted and the received signals over a noisy channel is formalized as

\begin{align}
\mathbf {W}=&\arg \min  \mathcal{E}\left|\left|\begin{bmatrix} \mathbf {H}_{[1]} \\  \mathbf {H}_{[2]} \\\vdots \\
\mathbf {H}_{[\rho]}
\end{bmatrix} \begin{bmatrix}\mathbf {W}
\end{bmatrix} \begin{bmatrix}\mathbf {s}
\end{bmatrix}+\begin{bmatrix}\mathbf {n}_{[1]} \\
\mathbf {n}_{[2]} \\
\vdots\\
\mathbf {n}_{[\rho]}
\end{bmatrix} -\begin{bmatrix}\mathbf {s} \\
\mathbf {s} \\
\vdots\\
\mathbf {s}
\end{bmatrix}\right|\right| ^{2} \notag\\
s. t. \ & \mathcal E|| \mathbf {Ws}||^{2} = P_{n}\label{eq: least squares extended SPC},
\end{align}
for the case that we need to serve $\rho = N_u/G$ users in each group using the same precoder.
Problem \eqref{eq: least squares extended SPC}
can be analytically solved, in the fashion of \cite{Choi2003},  by  noting that the cost function is the following sum:
\begin{align}
C_{\eqref{eq: least squares extended SPC}}= &\notag
  \mathrm{Tr}\left[(\mathbf {H}_{[1]}\mathbf{W-I})(\mathbf {H}_{[1]}\mathbf{W-I})^\dag\right] + \beta\mathrm{Tr}\notag
\left[ \mathbf{WW}^\dag\right]\\ &+ \dots\notag\\
&\mathrm{Tr}\left[(\mathbf {H}_{[\rho]}\mathbf{W-I})(\mathbf {H}_{[\rho]}\mathbf{W-I})^\dag\right] + \beta\mathrm{Tr}\notag
\left[ \mathbf{WW}^\dag\right] \\
=&\sum_{i=1}^{\rho}\mathrm{Tr}\left[(\mathbf {H}_{[i]}\mathbf{W-I})(\mathbf {H}_{[i]}\mathbf{W-I})^\dag\right] \notag\\
&+ \rho\beta\mathrm{Tr}\notag
\left[ \mathbf{WW}^\dag\right]   \label{eq: cost func2}\end{align}
where $\beta= \sigma^2/P_{n}$. By differentiation we get
\begin{align}
&\nabla_{\mathbf W}C(\mathbf W) = 0\Leftrightarrow\\
&\mathbf {W}\left(\sum_{i=1}^{\rho}\mathbf {H}_{[i]}^\dag\mathbf {H}_{[i]}+\rho\beta\mathbf {I}\right) = \sum_{i=1}^{\rho}\mathbf {H}_{[i]}^\dag,\label{eq: equivalent precoder1}
\end{align}
 Thus the general solution reads as
\begin{align}
\mathbf {W} =\left(\frac{1}{\rho}\sum_{i=1}^{\rho}\mathbf {H}_{[i]}^\dag\mathbf {H}_{[i]}+\beta\mathbf {I}\right)^{-1}\frac{1}{\rho}\sum_{i=1}^{\rho}\mathbf {H}_{[i]}^\dag&\label{eq: equivalent precoder2}
\end{align}
Following a different derivation methodology, this result was firstly reported in \cite{Silva2009}.

\textit{Remark 1:} Under the assumption of Rayleigh fading,  the elements of $\mathbf H$ are independent  zero mean complex Gaussian instances. Subsequently, due to the central limit theorem, as the number of users per group $\rho$ increases then the precoder will tend to zero:
  \begin{align} \lim_{\rho\rightarrow\infty }\frac{1}{\rho}\sum_{i=1}^{\rho}\mathbf {H}_{[i]} = 0.
  \end{align}
The implications of \textit{Remark 1} can be seen when the system dimensions grow large and the channel matrices tend to be modeled as zero mean random variables. The main result is that the system performance will degrade as the number of users per group increases. Assuming a fixed number of groups, the degradation as the number of users increases has only been examined hitherto via simulations \cite{Karipidis2008,Silva2009}. Herein, an analytical proof for this result has been provided. Moreover, remembering that $\rho = N_u / G$, for a fixed number of users the performance is expected to degrade as the number of groups increases. Since each user belongs to only one group, the maximum number of groups is bounded by $N_u$. Hence, the best performance is expected for a one user per group configuration. In other words, multicasting is expected to perform worst, in terms of precoding gain, over unicasting. An expected result, if one considers that in multicasting the degrees of freedom at the transmit side are reduced.

The above results provide a multicast aware $\mathrm{MMSE}$ solution for the calculation of the precoding matrix. However, the main drawback  of this solution is that it does not account for the practical per-antenna constraints. The simplest heuristic to  overcome this obstacle is to re-scale the solution so that the per-antenna constraints are not violated \cite{Taricco2014}.  Despite the fact that such an operation invalidates the MMSE optimality of the solution, it provides a low complexity heuristic method to design the precoder. Re-scaling is achieved by multiplying each line of the precoding matrix with the square root of the inverse level of power over satisfaction of the corresponding antenna.
\subsection{Multicast Multigroup Beamforming under $\mathrm{PACs}$}
\label{sec: optimal}
The MMSE solution described in the previous section is  one candidate linear precoding method. Another approach, namely the weighted max-min fair optimization of \cite{Christopoulos2014ICC,Christopoulos2014} is considered hereafter. The main benefit of this approach lies in the optimality of the solution as far as the $\mathrm{PAC}$s are concerned. The weighted max-min fair  problem with $\mathrm{PAC}$s  reads
as\begin{empheq}[box=\fbox]{align}
\mathcal{F:}\    \max_{\  t, \ \{\mathbf w_k \}_{k=1}^{G}}  &t& \notag\\
\mbox{subject to } & \frac{1}{\gamma_i}\frac{|\mathbf w_k^\dag \mathbf h_i|^2}{\sum_{l\neq k }^G |\mathbf w_l^\dag\mathbf h_i|^2+\sigma_i^2 }\geq t, &\label{const: F SINR}\\
&\forall i \in\mathcal{G}_k, k,l\in\{1\dots G\},\notag\\
 \text{and to }\ \ \ \ & \left[\sum_{k=1}^G  \mathbf w_k\mathbf w_k^\dag  \right]_{nn}  \leq P_n, \label{eq: max-min fair power const }\\%
 &\forall n\in \{1\dots N_{t}\},\notag
 \end{empheq}
 where $\mathbf w_k\in \mathbb{C}^{N_t}$ and $t \in \mathbb{R}^{+}$.  Problem $ \mathcal{F}$ receives as inputs the  $\mathrm{PAC}$ vector $\mathbf p = [P_1, P_2\dots P_{N_t}]$ and the target $\mathrm{SNIR}$s vector $\mathbf g = [\gamma_1,\gamma_2, \dots \gamma_{N_u}]$. Following the common  in the literature notation for ease of reference, the  optimal objective value of $ \mathcal{F}$ will be denoted as $t^*=\mathcal{F}(\mathbf g,  \mathbf p)$ and the associated optimal point as $\{\mathbf w_k^\mathcal{F}\ \}_{k=1}^{G}$. A common $\mathrm{SNIR}$  target between multiple users is a special case of the above general formulation. Of particular interest is the case where the users that belong in the same group share the same target i.e. $\gamma_i = \gamma_{k},\ \forall i \in\mathcal{G}_k, k\in\{1\dots G\}$, since the performance of all co-group users will be defined by the worst user in the group.
Towards solving this elaborate  problem via the means of convex optimization,  the principles of Semi-definite Relaxation ($\mathrm{SDR}$) and Gaussian randomization \cite{Luo2010} need to be combined with bisection. The detailed solution of this problem is given in \cite{Christopoulos2014ICC,Christopoulos2014}, and is omitted herein for shortness. Detailed discussions on the complexity of this method are also given therein. For the purposes of this work, it is only mentioned that the  complexity  is higher than the multicast aware heuristics, but it still remains polynomial in order. The exact determination of the trade-off between complexity increase and performance gains is left for future work. The optimally fair solutions are compared to  the MMSE based heuristics in the following.

\section{ Performance Evaluation \& Application} \label{sec: performance}

Extensive numerical results that exhibit the applicability of precoding in satellite communications are presented. To the end of providing accurate results, the simulation setup of \cite{Christopoulos2013AIAA} is employed. The simulation  parameters are summarized in  Tab. \ref{tab: simulation params}. The achievable spectral efficiency of the  $k$-th user is directly linked with its $\mathrm{SNIR}_k$ through the $\mathrm{DVB-S2}$ \cite{DVB_S2_standard} achievable spectral efficiency\footnote{More up-to-date $\mathrm{DVB-S2X}$ spectral efficiency mapping has been considered in \cite{Christopoulos2014TWCOM}.}. More importantly, to account for adaptive coding and modulation ($\mathrm{ACM}$) and the fact that a single modulation and coding scheme (MODCOD) is applied to each frame, the $\rho$ UTs that are simultaneously served by the same frame are assumed to be using the MODCOD corresponding to lowest $\mathrm{SNIR}$ value in the group. This consideration is inline with the common multicast consideration that the user with the lowest rate in each group will determine the performance of the group. The multibeam satellite antenna pattern has been provided in \cite{satnex,Christopoulos2013AIAA}. From the 245 beams used to cover Europe, the focus herein is on a cluster of 9 beams, as depicted in Fig. \ref{fig: CA}. This assumption is inline with future multi-gateway considerations, where precoding will be performed in each gateway separately \cite{Chatzinotas_GlobCom11}. Perfect channel state information is assumed throughout this work. The complex channel coefficients are generated as described in Sec. \ref{sec: Satellite Channel}, where only the phases due to different propagation paths between the satellite and users are assumed \cite{Zheng2012}.  Herein, the interferences from adjacent clusters are not accounted for, since the purpose is to give a relative comparison between the possible precoding methods rather than an absolute evaluation of  the total system throughput. For ease of reference, however,  the results are given on a per beam basis.
Finally, it needs to be stressed, that since the purpose of this work is to establish the most promising precoding method, no user scheduling is considered. Therefore, the results presented hereafter are averaged over uniformly random over the coverage distributed users, when a random schedule is considered in each frame.

The per beam achievable throughput with respect to an increasing on board available power budget for the conventional 4 color frequency reuse scheme and the two proposed precoding methods is given in Fig. \ref{fig: power1}. For a nominal on board power of 55~dBW, the weighted fair solution achieves 42\% improvement over the conventional system, while  the heuristic average precoder 21\%. In the same figure, the substantial gain of the proposed techniques with respect to an increasing power budget is also presented. This  is gain identical for both precoding methods. Fig.  \ref{fig: power2} presents the per beam throughput when four users per frame are considered. For this setting, the heuristic sub-optimal system performs worst than the conventional systems. However, the multicast approach still manages to achieve some gains (6\%).

To investigate the sensitivity of all methods to the frame dimensions, the per beam throughput is plotted with respect to an increasing number of users per frame in Fig.  \ref{fig: users}. The performance degradation of all precoding methods with the increasing number of users per frame is apparent. This expected result \cite{Christopoulos2014ICC} is justified by \textit{Remark 1.} This is intuitively expected by the inherent constraints of linear precoding methods. As the number of users increases, the transmit spatial degrees of freedom do not suffice to manage interferences and the performance is degraded. Nevertheless, the optimal multicast scheme manages to maintain gains over the conventional systems for up to five users per frame, whereas the heuristic scheme provides gains for up to two users per frame.

In Figs. \ref{fig:  rate1} and \ref{fig:  rate2}  the per user rate distribution over the coverage area for two and four users per frame respectively is plotted. In these figures, insights on the origins of the gains of the optimal multicast approach are gained. The fairness optimization, reduces the variability of the $\mathrm{SNIR}$ across the coverage area and consequently inside each frame. This results in better utilization of resources since  similar in terms of $\mathrm{SNIR}$s users are served by the same frame. On the contrary, the MMSE precoding approach exhibits high  $\mathrm{SNIR}$ variability. Hence, users with different $\mathrm{SNIR}$s are scheduled in the same frame and their performance is compromised by the performance of the worst user. Additionally,   many users are driven to the unavailability region, since their $\mathrm{SNIR}$ is lower than the minimum value that the available MODCODs can support. As depicted in Figs. \ref{fig:  rate1} and \ref{fig:  rate2}, with heuristic  MMSE precoding, more than 15\% and 30\% of users experience unavailability incidents over the coverage area respectively and therefore  receive zero rate.

\begin{figure}[h]
\centering
 \includegraphics[width=1\columnwidth]{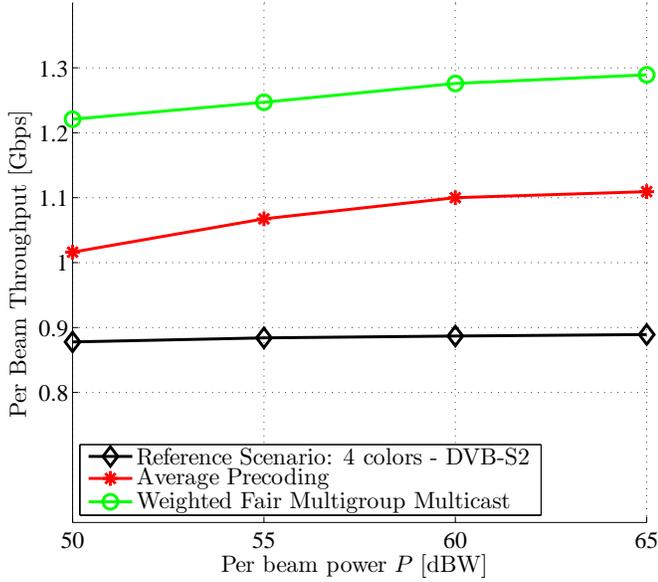}\\
 \caption{Per beam throughput performance versus increasing on board power for $\rho = 2$ users per frame.}
 \label{fig:  power1}
 \end{figure}
\begin{figure}[h]
\centering
 \includegraphics[width=1\columnwidth]{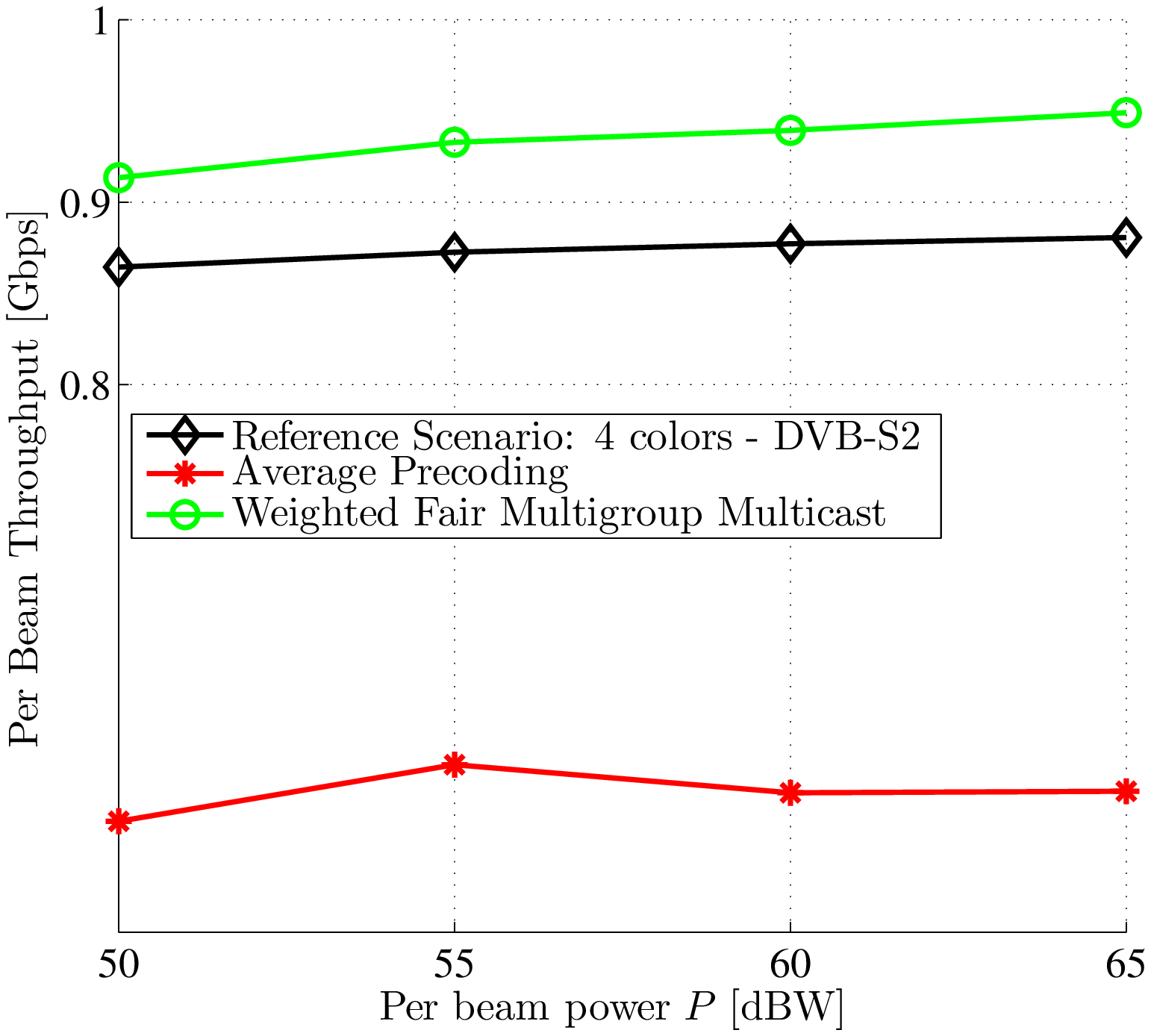}\\
 \caption{Per beam throughput performance versus increasing on board power for $\rho = 4$ users per frame.}\label{fig:  power2}
 \end{figure}
\begin{figure}[h]
\centering
 \includegraphics[width=1\columnwidth]{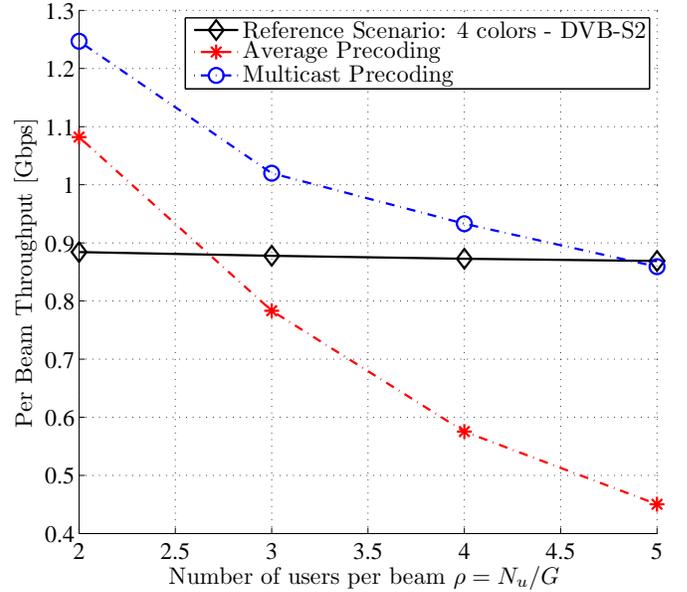}\\
 \caption{Per beam throughput versus  number of users per frame, for $P = 55$~dBW.}
 \label{fig: users}
 \end{figure}
\begin{figure}[h]
\centering
\includegraphics[width=1\columnwidth]{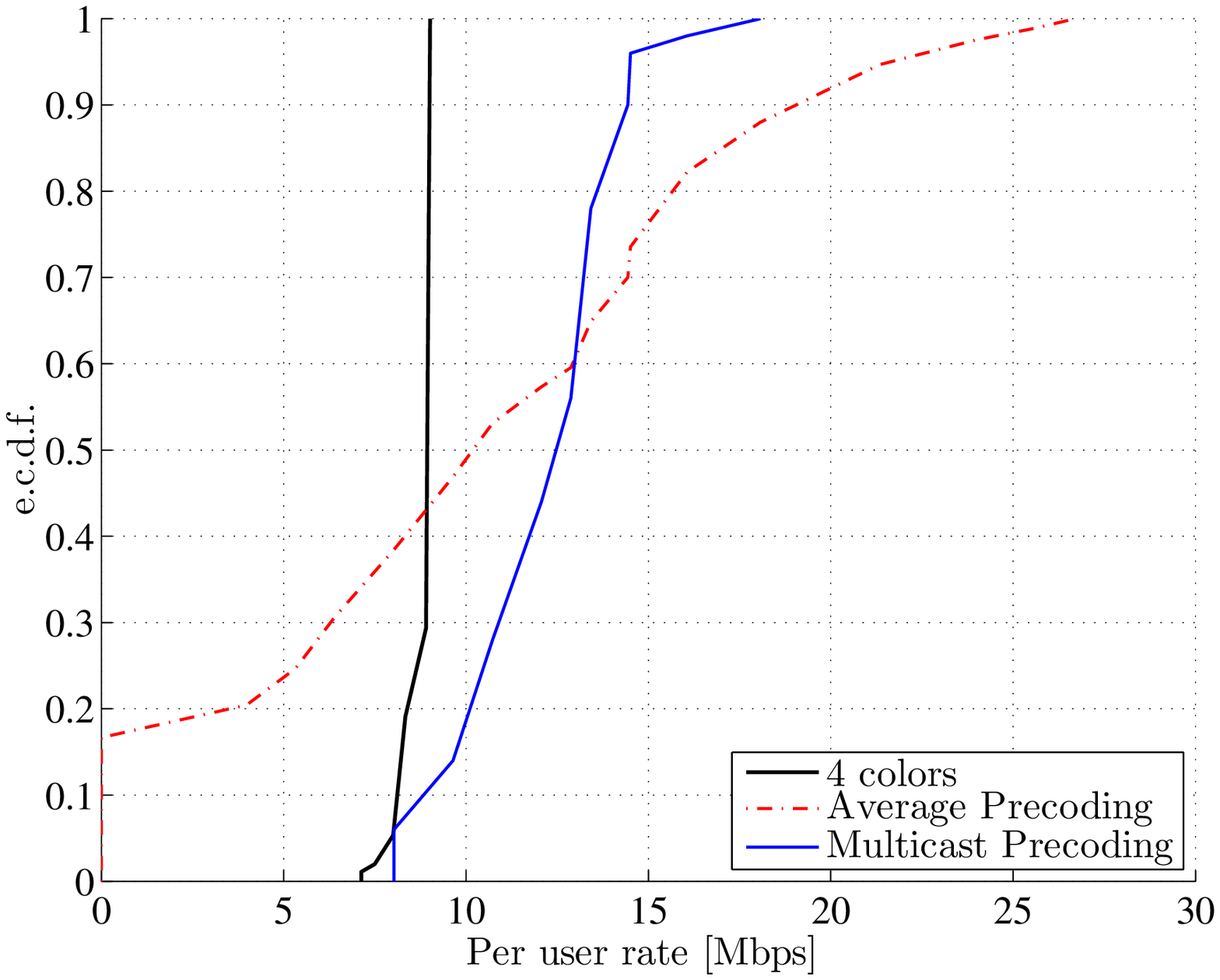}\\
\caption{Per user rate distribution over the coverage    for $P = 55$~dBW and  $\rho = 2$ users per frame. }
\label{fig:  rate1}
\end{figure}
\begin{figure}[h]
\centering
\includegraphics[width=1\columnwidth]{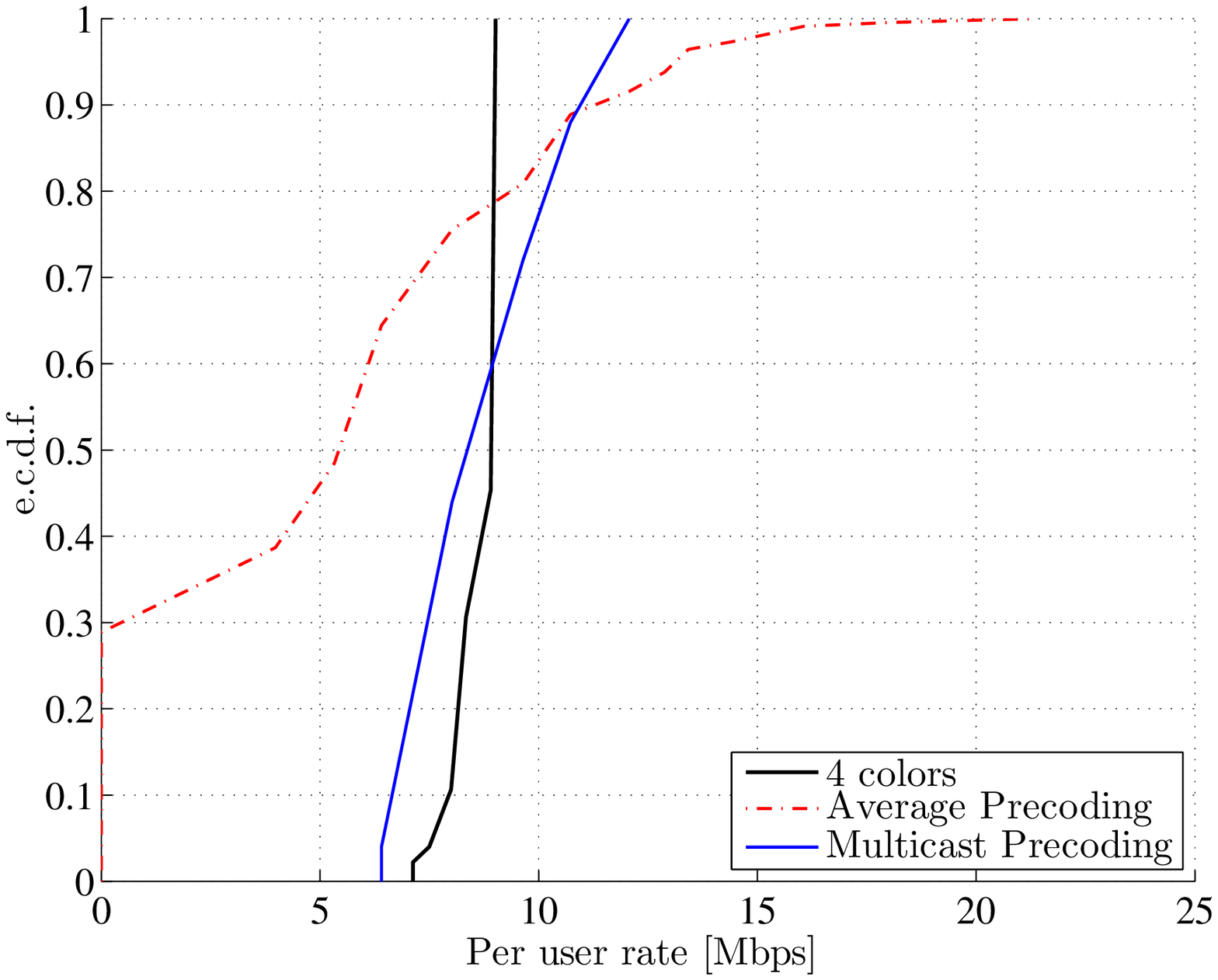}\\
\caption{Per user rate distribution over the coverage   for $P = 55$~dBW and $\rho = 4$ users per frame.}
\label{fig:  rate2}
\end{figure}

\section{Conclusions \& Future Work}\label{sec: conclusions}
In the present work, the optimal in a fairness sense, per-antenna power constrained, multigroup multicast linear precoding vectors have been applied in a multibeam satellite transmitter configured to operate in  full-frequency reuse. The throughput performance of the weighted fair multigroup multicast precoding is compared to the heuristic multicast aware MMSE solutions  re-scaled to respect the $\mathrm{PAC}$s. Simulation results over an accurate multibeam satellite scenario exhibit the superiority of the multigroup multicast solution over heuristic precoding methods. Insights on the origin of this result are provided. Finally, a sensitivity analysis with respect to system design parameters reveals the limits of the herein considered precoding methods.

Extensions of this work \cite{Christopoulos2014Globecom,Christopoulos2014TWCOM} include a sum-rate maximizing precoding design, adapted to the needs of satellite communications. Also,  the multiuser gains offered by proper user scheduling are gleaned in \cite{Christopoulos2014TWCOM} towards  establishing the applicability of precoding in satellite communications.
\section*{Acknowledgement}
This work was partially supported by the National Research Fund, Luxembourg under the CORE projects ``$\mathrm{CO^2SAT}$: Cooperative and Cognitive Architectures for Satellite Networks'' and ``$\mathrm{SEMIGOD}$: SpEctrum Management and Interference mitiGation in cOgnitive raDio satellite networks''. The authors would like to thank the work of the anonymous reviewers that greatly improved the presentation of this work. Acknowledgements are due to the European Space Agency and specifically Dr. P.-D. Arapoglou and Dr. A. Ginesi  for the definition of the frame based precoding problem in satellite communications.

\end{document}